\documentclass{llncs}

\usepackage{fullpage}
\usepackage{amsfonts,amssymb,amsmath}
\usepackage{wasysym}
\usepackage{stmaryrd}

\newcommand{\coh}{{\sf Coh}}
\newcommand{\adh}{{\sf Adh}}

\newcommand{\con}[1]{{#1}}

\newcommand{\aaa}{a}
\newcommand{\bbb}{b}
\renewcommand{\triangleright}{{\scriptscriptstyle\rhd}}
\renewcommand{\triangleleft}{{\scriptscriptstyle\lhd}}

\renewcommand{\blacktriangleright}{{\scriptscriptstyle\RHD}}
\renewcommand{\blacktriangleleft}{{\scriptscriptstyle\LHD}}
\newcommand{\blackbowtie}{{\blacktriangleright\!\!\blacktriangleleft}}

\newcommand{\AAA}{{\cal A}}

\newcommand{\EEE}{{\cal E}}

\newcommand{\NNN}{{\cal N}}

\newcounter{countroman}
{\begin{list}{{\rm (\roman{countroman})}}{\usecounter{countroman}}}%
{\end{list}}

\newcounter{countalpha}
{\begin{list}{(\alph{countalpha})}{\usecounter{countalpha}}}%
{\end{list}}

\newcounter{countalphabf}
{\protect\begin{list}{{\rm (}{\bf \protect\alph{countalphabf}}{\rm%
)}}{\protect\usecounter{countalphabf}}}% 
{\end{list}}

\newcommand{\bitmz}{\vspace{-.5\baselineskip}\begin{itemize}}
\newcommand{\eitmz}{\end{itemize}\vspace{-.25\baselineskip}}

\newcommand{\bdesc}{\vspace{-.5\baselineskip}\begin{description}}
\newcommand{\edesc}{\end{description}\vspace{-.25\baselineskip}}

\mathcode`\<="4268 %left delimiter
\mathcode`\>="5269 %right delimiter
\mathchardef\gt="313E %relation >
\mathchardef\lt="313C %relation <

\newcommand{\be}[1]{\begin{#1}}

\newcommand{\ee}[1]{\end{#1}}
\newcommand{\beq}{\begin{equation}}
\newcommand{\eeq}{\end{equation}}
\newcommand{\ba}[1]{\begin{array}{#1}}
\newcommand{\ea}{\end{array}}
\newcommand{\bea}{\begin{eqnarray}}
\newcommand{\eea}{\end{eqnarray}}
\newcommand{\bear}{\begin{eqnarray*}}
\newcommand{\eear}{\end{eqnarray*}}
\newcommand{\bpr}{\begin{prf}{}}
\newcommand{\epr}{\end{prf}}
\newcommand{\bprf}[1]{\begin{prf}{#1}}
\newcommand{\eprf}{\end{prf}}

\renewcommand{\to}{\xymatrix@C-.5pc{\ar[r]&}}
\newcommand{\ot}{\xymatrix@C-.5pc{& \ar[l]}}
\newcommand{\tto}[1]{\xymatrix@C-.5pc{\ar[r]^{#1}&}}
\newcommand{\oot}[1]{\xymatrix@C-.5pc{&\ar[l]_{#1}}}

\newcommand{\mono}{\xymatrix@C-.5pc{\ar@{>->}[r]&}} 
\newcommand{\epi}{\xymatrix@C-.5pc{\ar@{->>}[r]&}}
\newcommand{\mmono}[1]{\xymatrix@C-.5pc{\ar@{>->}[r]^{#1}&}} 
\newcommand{\eepi}[1]{\xymatrix@C-.5pc{\ar@{->>}[r]^{#1}&}}
\renewcommand{\mapsto}{\xymatrix@C-.5pc{\ar@{|->}[r]&}}
\newcommand{\mmapsto}[1]{\xymatrix@C-.5pc{\ar@{|->}[r]^{#1}&}}
\newcommand{\inclusion}{\xymatrix@C-.5pc{\ar@{^{(}->}[r] &}}
\newcommand{\iinclusion}[1]{\xymatrix@C-.5pc{\ar@{^{(}->}[r]^{#1}&}}
\newcommand{\dtto}[2]{\xymatrix@C-.5pc{\ar@<.875mm>[r]^{#1} \ar@<-.875mm>[r]_{#2}&}}

\newcommand{\draw}{capacity }

\newcommand{\path}{path }
\newcommand{\Path}{Path }

\newcommand{\cn}[1]{\tilde{#1}}
\newcommand{\pn}[1]{\widehat{#1}}
\newcommand{\pr}[1]{{#1}}

\newcommand{\bias}{\Upsilon}

\usepackage{xy} 
\xyoption{all} 
\CompileMatrices 

\title{
Network as a computer:\\ 
ranking paths to find flows
}
\author{Dusko Pavlovic\thanks{Email: {\tt dusko@\{kestrel.edu,comlab.ox.ac.uk\}}. 
Supported by EPSRC and ONR.}\inst{}}
\institute{Oxford University and Kestrel Institute}
\date{}

\begin{document} 
\maketitle

\begin{abstract} 
We explore a simple mathematical model of network computation, based on Markov chains. Similar models apply to a broad range of computational phenomena, arising in networks of computers, as well as in genetic, and neural nets, in social networks, and so on. The main problem of interaction with such spontaneously evolving computational systems is that the data are not uniformly structured. An interesting approach is to try to extract the semantical content of the data from their distribution among the nodes. A concept is then identified by finding the community of nodes that share it. The task of data structuring is thus reduced to the task of finding the network communities, as groups of nodes that together perform some non-local data processing. Towards this goal, we extend the ranking methods from nodes to paths, which allows us to extract information about the likely flow biases from the available static information about the network.
\end{abstract}

\section{Introduction}\label{Introduction}
Initially, Web search was developed as an instance of {\em information retrieval}, optimized for a particularly large distributed database. With the advent of online advertising, Web search got enhanced by a broad range of {\em information supply\/} techniques where the search results are expanded by additional data, extrapolated from user's interests, and from search engine's stock of information. From the simple idea to match and coordinate the push and the pull of information on the Web as a new computational platform \cite{OReilly:Web2} sprang up a new generation of web businesses and social networks. Similar patterns of information processing are found in many other evolutionary systems, from gene regulation, protein interaction and neural nets, through the various networks of computers and devices, to the complex social and market structures \cite{Newman:Networks}.

This paper explores some simple mathematical consequences of the observation that the Web, and similar networks, are much more than mere information repositories: besides storing, and retrieving, and supplying information, they also generate, and process information. We pursue the idea that the Web can be modeled as a computer, rather than a database; or more precisely, as a vast % information processing plant
multi-party computation \cite{Goldreich2}, akin to a market place, where masses of selfish agents jointly evaluate and generate public information, driven by their private utilities. While this view raises interesting new problems across the whole gamut of Computer Science, the most effective solutions, so far, of the problem of {\em semantical\/} interactions with the Web computations were obtained by rediscovering and adopting the {\em ranking\/} methods, deeply rooted in the sociometric tradition \cite{Katz,Hubbell}, and adapting them for use on very large indices, leading to the whole new paradigm of {\em search\/} \cite{page98pagerank,kleinberg99authoritative,Langville06google}. Implicitly, the idea of the Web as a computer is tacitly present already in this paradigm, in the sense that the search rankings are extracted from the link structure, and other intrinsic information, generated on the Web itself, rather than stored in it. 

\paragraph{Outline of the paper.} In section \ref{Basic} we introduce the basic network model, and describe a first attempt to extract information about the flows through a network from the available static data about it. In sections \ref{Adding paths} and \ref{Dynamics}, we describe the structure which allows us to lift the notion of rank, described in section \ref{Rank}, to path networks in section \ref{Path networks}. Ranking paths allows us to extract a random variable, called attraction bias, which allows measuring the {\em mutual information\/} of the distributions of the inputs and the outputs of the network computation, which can be viewed as an indicator of non-local information processing that takes place in the given network. In the final section, we describe how the obtained data can be used to detect semantical structures in a network. The experimental work necessary to test the practical effectiveness of the approach is left for future work.

\section{Networks}\label{Basic}
\subsubsection*{Basic model.} We view a network as an edge-labelled directed graph 
$
A   =   \big(R \oot{\gamma} E \dtto{\delta}{\varrho} N\big)
$, where $N$ and $E$ are, respectively, the finite sets of {\em nodes}, 
and  {\em links}, or {\em edges}, whereas $R$ is an ordered field of {\em values} (in some applications an ordered ring of functions). A link $i\stackrel{e}{\rightarrow} j$ is thus an element $e\in E$ with $\delta(e) = i$ and $
\varrho(e)=j$. The value $\gamma(e)$ is the cost (when positive), or 
payoff (when negative) of the traffic over $e$. These data induce the 
{\em adjacency matrix\/} $E = (E_{ij})_{N\times N}$ and the {\em 
{\draw} matrix\/} $A = (A_{ij})_{N\times N}$, with the entries
%\[
%E_{ij}\ = \  \{e \in E\ |\  i\stackrel{e}{\rightarrow} j \} \qquad \qquad
%A_{ij} \ = \ \sum_{e\in E_{ij}} A_e
%\]
$
E_{ij} =  \{e \in E\ |\  i\stackrel{e}{\rightarrow} j \}$ and 
$A_{ij} =  \sum_{e\in E_{ij}} A_e$, 
where $A_e =  2^{-\gamma(e)}$ is the capacity of the link $e$. 

\paragraph{Remark.} The term "capacity" is used here as in network flow theory.\footnote{The information theoretic homonym  has a different, albeit related meaning, which motivates the choice of $\gamma(e) = -\log_2 A_e$.} The cost or the payoff of a link may represent its value in a pay-per-click model of a fragment of the Web; or it may denote the proximity of the web pages within the same site, or within a group of interconnected sites. In a protein network, the energy cost or payoff may be derived from the chemical affinities between the nodes. While this parameter can be abstracted away, simply by taking $\gamma(e)=0$ for all links $e$, its role will become clear in sections \ref{Adding paths} and \ref{Path networks}, where it allows discounting and eliminating some paths.

\subsubsection*{Basic dynamics.} The simplest model of network dynamics is based on the assumption that the traffic flows are distributed proportionally to the link capacities. Randomly sampling the Web traffic, we shall thus find a 
surfer on a link $e$ with the probability $\alpha_e = \frac{A_e}{A_
\bullet}$, where $A_\bullet = \sum_{f\in E} A_f$. 

In order to find the communities in a network, we need to detect the 
traffic biases between its nodes. We assume that the traffic between 
the nodes within the same community will be higher than the capacity 
of the links between them would lead us to expect; and that the traffic 
between the different communities will be lower 
than expected. To measure such traffic biases, we normalize the \draw matrix $A$ to get the {\em \draw distribution\/} $\alpha = (\alpha_{ij})_
{N\times N}$ as
$
\alpha_{ij}  =  \frac{A_{ij}}{A_{\bullet\bullet}}
$,
where $A_{\bullet\bullet} = \sum_{k,\ell\in N} A_{k\ell}$. The entry $
\alpha_{ij}$ is thus the probability that  a random sample of traffic on 
$A$, following the simple dynamics proportional to capacity, will be 
found on a link from $i$ to $j$. On the other hand, the marginals of the 
probability distribution $\alpha$,
\[
\alpha_{i\bullet}\ = \ \sum_{j\in N} \alpha_{ij}\qquad \qquad
\alpha_{\bullet j}\ = \  \sum_{i\in N} \alpha_{ij}
\]
correspond, respectively, to the probabilities that a random sample of 
traffic will have $i$ as its source, and $j$ as its destination. Let us call $\alpha_{i\bullet}$ the {\em out-rank\/} of $i$, and $\alpha_{\bullet j}$ the {\em in-rank\/} of $j$, because they can be viewed as the simplest, 
albeit degenerate cases of the notion of rank. 

If the in-rank and the out-rank are statistically independent, then (by the 
definition of independence) the probability that a random traffic sample 
goes from $i$ to $j$ will be $\alpha_{i\bullet} \alpha_{\bullet j}$. Their 
dependency is thus measured by the {\em traffic bias} matrix $\upsilon 
=(\upsilon_{ij})_{N\times N}$ with the entries
$
\upsilon_{ij}  =  \alpha_{ij} - \alpha_{i\bullet}\alpha_{\bullet j}
$
falling in the interval $[-1,1]$. The higher the bias, the more 
unexpected traffic there is. For a set of nodes $U\subseteq N$ the 
values
\[
\coh(U)\ =\ \sum_{i,j \in U} \upsilon_{ij} \qquad \qquad
\adh(U)\ =\ {\sum_{i\in U,j\not\in U}} \upsilon_{ij} + \upsilon_{ji}
\]
can thus be construed as the {\em cohesion\/} and the {\em adhesion
\/} forces: the total traffic bias within the group, and with its exterior, 
respectively. A network community $U$ can thus be recognized as a 
set of nodes with a high cohesion and a low adhesion \cite
{newman06modularity}. The idea that semantically related nodes can be captured as members of the same network communities, derived from the graph structure, is a natural extension of the ranking approach, which has been formalized in  \cite{Haveliwala,Ollivier-Senellart}. 

The only problem with applying that idea to the above model is that our initial assumption --- that the traffic distribution on $A$ is proportional to its link capacities --- is not very realistic. It abstracts away all traffic dynamics. On the other hand, the static network model, as given above, does not provide any data about the actual traffic. We explore the ways to solve this problem, and extract increasingly more realistic views of traffic dynamics from a static network model.

\section{Adding paths}\label{Adding paths}
A path $i\stackrel{a}{\rightarrow} j$ in a network $A$ is a sequence of 
links $i \stackrel{a_1}{\rightarrow} k_1 \stackrel{a_2}{\rightarrow}k_2
\rightarrow \cdots \stackrel{a_n}{\rightarrow} j$. In many cases of 
interest, traffic dynamics on a network depends on the path selections, 
rather than just on single links.

One idea is to add the paths to the structure of a network, and to 
annotate how the links compose into paths, and how the paths 
compose into longer paths. This amounts to generating the free 
category \cite{MacLane} over the network graph. 
Unfortunately, adding all paths to a network usually  destroys some 
essential information, just like the transitive closure of a relation does. 
E.g., in a social network, a friend of a friend is often not even an 
acquaintance. Taking the transitive closure of the friendship relation 
obliterates that fact. Moreover, the popular "small world" phenomenon 
suggests that almost {\em every two people\/} can be related through 
no more than six friends of friends of friends\ldots So already adding 
all paths of length six to a social network, with a symmetric friendship 
relation, is likely to generate a complete graph. In fact, the average 
probability that two of node's neighbors in an undirected graph are 
also linked with each other is an important factor, called {\em 
clustering coefficient\/} \cite{Watts-Strogatz}. On the other hand, in 
some networks, e.g. of protein interactions, a link $i\rightarrow k$ 
which shortcuts the links $i\rightarrow j\rightarrow k$ often denotes a 
direct {\em feed-forward} connection, rather than a composition of the 
two links, and leads to essentially different dynamics.

So only "short" paths must be added to a network: composition must 
be penalized.

\be{definition}
For a given network $A   =  \big(R \oot{\gamma} E \dtto{\delta}{\varrho} 
N\big)$, a cutoff value $v\in R$, and a composition penalty $d\in R$, 
we define the {\em  $v$-completion\/} to be the network
$
A^{*v}\   = \  \big(R \oot{\gamma} E^{*v} \dtto{\delta}{\varrho} N\big)
$,
where\\ $E^{*v}\ =\  \{a \in E^*\ |\ \gamma(a)\leq v\}$ and 
\bear
%E^{*v} & = & \{a \in E^*\ |\ \gamma(a)\leq v\}\\ 
\gamma\big(i_0 \stackrel{a_1}{\rightarrow} i_1 \stackrel{a_2}
{\rightarrow}i_2\rightarrow \cdots  \stackrel{a_n}{\rightarrow} i_n\big) & 
= & (n-1)d + \gamma(a_1) +\cdots+ \gamma(a_n)
\eear
and $E^*$ is the set of all nonempty paths in $A$.
\ee{definition}

\paragraph{Remarks.} $E^*$ can be obtained as the matrix of sets 
$E^* = \sum_{n=0}^\infty E^n$ where each $E^n$ is a power of the 
adjacency matrix $E$. If the entry $E_{ij}$ is viewed as the set of links 
$\{i\stackrel{e}\rightarrow j\}$, then the entry $E^2_{ij} = \sum_{k=1}^N\ 
E_{ik}\cdot E_{kj}$ corresponds to the set of 2-hop paths $\{i\stackrel
{e_1}\rightarrow k \stackrel{e_2}\rightarrow j\}$ through the various 
nodes $k$; the matrix $E^3$ similarly corresponds to the matrix of 3-hop paths, and so on. 

The $v$-closed network $A^{*v}$ is closed under the composition of 
low cost paths, but not if the cost is greater than $v$. It is not hard to 
see that the $v$-completion is an idempotent operation, i.e. $A^{*v*v} 
= A^{*v}$, but it may fail to be a proper closure operation, because a 
link $e$ in $A$, with $\gamma(e)\gt v$, may lead to $A \not\subseteq 
A^{*v}$.

In the rest of the paper, we assume that the networks are $v$-
complete for some $v$, i.e. $A = A^{*v}$. This means that the relevant 
pathways are already represented as links, with the composition 
penalty absorbed in the cost.

\section{Network dynamics}\label{Dynamics}
In order to derive network dynamics from a static network model, one 
first specifies the way in which the behavior of a computational agent, 
processing data on the network, is influenced by the network structure, 
and then usually derives a Markov chain that drives the traffic. The network features that influence its dynamics can then be incrementally 
refined, yielding more and more information. 

\subsection{Forward and backward}
Random walks on networks are often represented in terms of the behavior of surfers on the Web, following the hyperlinks.\footnote{The 
surfers deserve their name by following the "waves", i.e. obeying the 
same dynamics.} The simplest surfer behavior chooses an out-link uniformly at random at each node. A visitor of a node $i$ 
will thus proceed to a node $j$ with probability 
$
A^\triangleright_{ij} \ =\  \frac{A_{ij}}{A_{i\bullet}}
$,
where $A_{i \bullet } =  \sum_{k=1}^N A_{ik}$ is the out-degree of $i$. 
The row-stochastic matrix $A^\triangleright = (A^\triangleright_{ij})_{N
\times N}$ represents {\em forward dynamics\/} of a network $A$. The 
entries $A^\triangleright_{ij}$ are called the {\em pull\/} coefficients of 
$i$ by $j$.

Dually, {\em backward dynamics\/} of a network $A$  is represented by 
a column-stochastic matrix $A^\triangleleft= (A^\triangleleft_{ij})_{N
\times N}$, where the entry 
$
A^\triangleleft_{ij}\ =\  \frac{A_{ij}}{A_{\bullet j}}
$,
with $A_{\bullet j} =  \sum_{k=1}^N A_{kj}$ denoting the in-degree of $j$,  describes the probability that a surfer who is on the node $j$ came 
there from the node $i$. The entries $A^\triangleleft_{ij}$ are called the 
{\em push\/} coefficients.

\paragraph{Remark.} Note that the capacity matrix can be normalized to get $A^\triangleright$ and $A^\triangleleft$ as above only if no rows, resp. 
columns, consist of 0s alone. This means that every node of the network must have at least one out-link, resp. at least one in-link. Networks that do not satisfy this requirement need to be modified, in one way or 
another, in order to enable analysis. Adding a high-cost link between 
every two nodes is clearly the minimal perturbation (with maximal 
entropy) that achieves this.  Alternatively, the problem can also be 
resolved by adjoining a fresh node, and the high-cost links in and out 
of it \cite{Bianchini05inside}. Either way, the quantitative effect of such 
modifications can be made arbitrarily small by increasing the cost of 
the added links.

\subsection{Forward-out and backward-in dynamics}
The next example can be interpreted in two ways, either to show how forward and backward dynamics can be refined to take into account various navigation capabilities, or  how to abstract away irrelevant cycles. Suppose that a surfer searches for the hubs on the network: he prefers to follow the hyperlinks that lead to the nodes with a higher out-degree. This preference may be realized by annotating the hyperlinks according to the out-rank of their target nodes. Alternatively, the surfer may explore the hyperlinks ahead, and select those with the highest out-degree; but we want to ignore the exploration part, and simply assume that he proceeds according to the out-rank of the 
nodes ahead. The probability that this surfer will move from $i$ to $j$ is thus
\[
\con{A}^\blacktriangleright_{ij}\ \ =\ \  \con{A}^\triangleright_{ij} \cdot 
\con{\pr{\alpha}}_{j\bullet}\ \ = \ \ \frac{\con{A}_{ij} \con{\pr{A}}_{j\bullet}}
{\con{A}_{i\bullet}\con{\pr{A}}_{\bullet\bullet}}
\]  
We call this the {\em forward-out\/} dynamics. In the dual, {\em 
backward-in\/} dynamics, the surfers are more likely to arrive to $j$ 
from $i$ if this is a frequently visited node, i.e. if its in-rank is higher
\[
\con{A}^\blacktriangleleft_{ij} \ \ =\ \  \con{\pr{\alpha}}_{\bullet i}\cdot 
\con{A}^\triangleleft_{ij}\ \ =\  \  \frac{\con{\pr{A}}_{\bullet i} \con{A}_{ij}}{ 
\con{\pr{A}}_{\bullet\bullet} \con{A}_{\bullet j}}
\]
These dynamics will be the particularly convenient to demonstrate an example of bias analysis in section \ref{Path networks}, because they clearly display clearly how the simple traffic bias $\upsilon$ from section \ref{Basic} can be refined by the various dynamics factors.

\subsection{Teleportation and preference}
The main point of formulating network dynamics, especially in the Markov chain form, is to be able to compute the node ranking as its invariant distribution. However, since the network graphs are usually {\em not\/} strongly connected, the Markov chains, derived from their structure, are often reducible to classes of nodes with no way out.

The simplest remedy is the idea of {\em teleportation}, going back to \cite{page98pagerank}. A general interpretation is that, whichever dynamics a surfer might follow, at each node he tosses a biased coin, and with a probability $d\in (0,1)$ follows that dynamics. Otherwise, with a probability $1-d$, he "teleports" to a randomly chosen node, ignoring all hyperlinks and other structure. Following, say, forward dynamics, the probability that he will go from $i$ to $j$ is thus
$
A^P_{ij} \ = \ dA^\triangleright_{ij} + \frac{1-d}{N}
$. 
This is roughly the PageRank dynamics, from which the Google 
search engine had started \cite{page98pagerank}\footnote{The original version allowed $A^\triangleright_{ij}$ to be 0, if $A_{i\bullet}$ is 0, i.e. if $i$ is a "sink-hole", and the teleportation factor was added to save dynamics from such sinkholes. Other modifications were introduced later.}. The induced dynamics is thus
$
A^P \ = \ dA^\triangleright + (1-d)P
$, 
where $P = (P_{ij})_{N\times N}$ has all entries $P_{ij}= \frac{1}{N}$. In the networks without a cost function, this is interpreted as adding a link between every two nodes. The influence of such links can be controlled using the cost functions. In any case, the resulting Markov chains become irreducible, and their stationary distributions do not get captured in any closed components. Furthermore, the model can be {\em personalized\/} by capturing surfer's preferences in terms of the biases in $P$: the entries $P_{ij}$  can be interpreted as $i$'s  {\em trust\/} in $j$ \cite{trustrank}.  The extensions of the {\em backward\/} dynamics by teleportation yields to different interpretations, which the reader may wish to consider on her own.

\section{Ranking}\label{Rank}
Intuitively, the rank of a node is the probability that randomly 
sampled traffic will be found to visit that node. In search, this is taken as a generic relevance measure. The technical implication is that the rank can be obtained as a stationary distribution of the Markov chain capturing dynamics. Each notion of dynamics thus induces a corresponding notion of rank. Since a Markov chain can 
be viewed as a linear, and hence continuous transformation of the 
simplex of distributions, which is closed and compact, already 
Brouwer's fixed point theorem guarantees that the rank always exists. Finding a meaningful, useful notion of rank is another matter.

First of all, as already mentioned, networks often decompose into loosely connected subnets. In the long run, all traffic is likely to get captured in some such subnet. This results in multiple stationary distributions, each concentrated in a closed subnet, zero otherwise. Dynamics derived directly from the network graph therefore result in uninformative ranking data. In order to assure that the relevant Markov chains are irreducible and aperiodic, and thus induce unique and nondegenerate stationary distributions, network dynamics usually need to be perturbed, using a damping and stabilizing factor such as teleportation. Another sort of problems arise when the unique stationary distribution is not an attractor, or when the rate of convergence is unfeasibly slow \cite{Golub03arnoldi,boldi05damping}. 

While very important in concrete applications, these problems, 
and their solutions, have less impact on the conceptual analyses pursued in this paper. {\em We shall henceforth assume that all processes have been adjusted to induce  unique and effectively computable ranking.}\footnote{This implies that all notions of dynamics that we consider have a tacit damping factor. We do not display it only because it needlessly complicates formulas.}   

\subsection{Promotion and reputation}
We now explain the intuition behind the simplest notions of rank.

In social terms, the push coefficient $A^\triangleleft_{ij} = \frac{A_{ij}}
{A_{\bullet j}}$ can be interpreted as measuring how much $i$ 
supports (or advocates) $j$. The concept of {\em promotion\/} can then 
be formalized as a probability distribution $r^\triangleleft$, such that 
$
r^\triangleleft_i \ = \ \sum_{k=1}^N A^\triangleleft_{ik} r^\triangleleft_k
$. 
In words, the promotion rank  (or {\em push rank\/}) $r^\triangleleft_i$ 
of a node $i$ is the sum of the promotion ranks $r^\triangleleft_k$ of 
its children nodes, each allocated to $i$ according to the push 
coefficient $A^\triangleleft_{ik}$, measuring $i$'s support for $k$.

Dually, the pull coefficient $A^\triangleright_{ij}$ can be interpreted as 
measuring how much $i$ trusts $j$. The concept of {\em reputation\/} 
can then be formalized as a probability distribution $r^\triangleright$, 
such that 
$
r^\triangleright_j\ = \ \sum_{k=1}^N r^\triangleright_k A^
\triangleright_{kj}
$. 
This reputation rank (or {\em pull rank\/}) $r^\triangleright_i$ of a node 
$i$ is thus the sum of the reputation ranks $r^\triangleright_k$ of its 
parent nodes, each allocated according to the pull coefficient $A^
\triangleright_{kj}$, of $k$'s trust in $j$.

Gathering the promotion values in a column vector $r^\triangleleft$ 
and the reputation values in a row vector $r^\triangleright$, we can 
rewrite the definitions of $r^\triangleleft$ and $r^\triangleright$ in the matrix form
\[
r^\triangleleft \ = \ A^\triangleleft r^\triangleleft  \qquad \qquad
r^\triangleright \ = \ r^\triangleright A^\triangleright
\]
The refined notions of promotion $\con{r}^\blacktriangleleft$ and 
reputation $\con{r}^\blacktriangleright$ are defined and interpreted 
along the same lines, as the stationary distributions of the  processes 
$\con{A}^\blacktriangleleft$ and $\con{A}^\blacktriangleright$ 
respectively.

\subsection{Expected flow}\label{Expected flow}
While dynamics of reputation has been studied for a long time \cite{Katz,Hubbell}, and with increased attention recently, since it become 
a crucial tool of Web search \cite{page98pagerank,Langville06google}, the dual dynamics of promotion does not seem to have attracted much attention. We need both notions to define the expected traffic flow.

The expected flow from $j$ to $k$, under the assumption that they are 
independent, is caused only by a "traffic pressure", resulting from the 
pull to $j$ and the push from $k$. Following this idea, we define 
\bea\label{diamond}
r^{\blackbowtie}_{jk} & = & r^\blacktriangleright_j r^\blacktriangleleft_k 
\eea 
The expected flow $r^\blackbowtie$ is thus a probability distribution 
over $N\times N$, which can be represented as the matrix $r^
\blackbowtie = r^\blacktriangleleft \cdot r^\blacktriangleright$, obtained 
by multiplying the column vector $r^\blacktriangleleft$ and the row 
vector $r^\blacktriangleright$. Since $r^\blacktriangleleft$ and $r^
\blacktriangleright$ are the principal eigenvectors of $A^
\blacktriangleleft$ and $A^\blacktriangleright$, $r^\blackbowtie$ is the 
unique distribution satisfying $r^\blackbowtie = A^\blacktriangleleft 
\cdot r^\blackbowtie \cdot A^\blacktriangleright$, i.e.
$r^{\blackbowtie}_{jk}\ =\  \sum_{i=1}^N \sum_{\ell=1}^N  A^
\blacktriangleright_{ij} r^{\blackbowtie}_{i\ell} A^\blacktriangleleft_{k\ell}$.
Intuitively, this means that the flow pressure from $i$ to $\ell$ propagates to cause a flow pressure from $j$ to $k$ proportionally to the force of the traffic from $i$ to $j$ and to the force of traffic flows from $k$ to $\ell$ --- {\em provided\/} that $j$ and $k$ are independent. In order to measure their dependency, we attempt to capture how the {\em 
actual flows}\/ from $i$ to $\ell$ (rather than mere flow pressure) may 
get diverted, say by the high costs and the low capacities, to cause 
actual flows from $j$ to $k$.

\section{\Path networks}\label{Path networks}
\begin{definition}\label{def2}
Given a $v$-closed network $A  =  \big(R \oot{\gamma} E \dtto{\delta}
{\varrho} N\big)$, we define the {\em \path network}\\ 
$
\pn{A}\  =\   \big(R \oot{\gamma} \pn{E} \dtto{\delta}{\varrho} \pn{N}
\big)
$, 
where 
$
\pn{N} \ = \ \pr{E}$, and
$
\pn{E} \ = \ \sum_{\aaa,\bbb\in E} \pn{E}_{\aaa\bbb}$, with
%$
%\pn{E}_{\aaa\bbb} \ = \  \big\{f=<f_0,f_1>\in E_{ij}\times E_{k\ell}\ |\  
%\notag \\ && \hspace{1em} \gamma(f_0)+ \gamma(\bbb)+ \gamma(f_1) 
%- \gamma(\aaa) \leq v-2d \big\}$, and
\bea
\pn{E}_{\aaa\bbb} & = &  \big\{f=<f_0,f_1>\in E_{ij}\times E_{k\ell}\ |\  
 \gamma(f_0)+ \gamma(\bbb)+ \gamma(f_1) - \gamma(\aaa) \leq v-2d \big\} 
 \\
\gamma(f) & = & 2d + \gamma(f_0) + \gamma(\bbb) + \gamma(f_1) - 
\gamma(\aaa) \label{gamma}
\eea
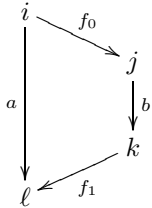
\[
\xymatrix@C-1pc@R-1.5pc
{ i \ar
[dddd]_\aaa 
\ar[drr]^{f_0}\\ 
 && j \ar
 [dd]^\bbb  \\  \\
&& k \ar[dll]^{f_1} \\ 
\ell 
}
\]
\end{definition}

\subsubsection{Dynamics of path selection.}
Recalling that $\pn{A}_{\aaa\bbb} = \sum_{f\in \pn{E}_{\aaa\bbb}} 2^{-
\gamma(f)}$, we define the forward and the backward dynamics, and 
the pull rank and the push rank just like before:
\begin{align*}
\pn{A}^\triangleright_{\aaa\bbb} & =  \frac{\pn{A}_{\aaa\bbb}}{\pn{A}_
{\aaa\bullet}}  & \pn{A}^\triangleleft_{\aaa\bbb} & =  \frac{\pn{A}_{\aaa
\bbb}}{\pn{A}_{\bullet \bbb}} \\
\pn{r}_\bbb & =  \sum_{\aaa\in \pn{N}} \pn{r}_\aaa\pn{A}^\triangleright_
{\aaa\bbb} &
\pn{r}^\triangleleft_\aaa & =  \sum_{\bbb\in \pn{N}} \pn{A}^\triangleleft_
{\aaa\bbb} \pn{r}^\triangleleft_\bbb
\end{align*}
Intuitively,  $\pn{A}^\triangleright_{\aaa\bbb}$ is now the probability 
that traffic through $\aaa$  is diverted to $\bbb$ (rather than to some 
other path); while $\pn{A}^\triangleleft_{\aaa\bbb}$ is the probability 
that traffic through $\bbb$ is diverted from $\aaa$ (and not from some 
other path). The pull rank $\pn{r}_\bbb$, i.e. the probability that $\bbb$ 
will be traversed,  can thus be understood as its {\em attraction}; 
whereas $\pn{r}^\triangleleft_\aaa$ is the probability that $\aaa$ will be 
avoided. 

Using the pull rank of the paths, we can now define the {\em node 
attraction\/} between $j$ and $k$ to be the total attraction of all paths 
between them:
\bea\label{natt}
\pn{r}_{jk} & = & \sum_{j\underset{\bbb}{\rightarrow}k} \pn{r}_\bbb
\eea
The idea is that this notion of attraction the nodes will allow us to refine 
the estimate of the traffic bias $\upsilon$ as described in section \ref{Basic}. In particular, 
consider {\em attraction bias}
\bea 
\bias_{jk} &= & \pn{r}_{jk} - r^{\blackbowtie}_{jk} \label{abias}
\eea
To motivate this, note that expanding the formula for $r^{\blackbowtie}_{jk}$ in section \ref{Expected flow} shows 
that $r^{\blackbowtie}$ is the stationary distribution  of the Markov 
chain ${A}^{\blackbowtie} = \left({A}^{\blackbowtie}_{(ij)(k\ell)}\right)_
{N^2\times N^2}$, where
\[
{A}^{\blackbowtie}_{(ij)(k\ell)}\ =\  \frac{A_{ij} \pr{A}_{j\bullet} \pr{A}_
{\bullet k} A_{k\ell}}{A_{i\bullet } \pr{A}^2_{\bullet\bullet} A_{\bullet\ell}}\qquad \mbox{ and } \qquad
r^{\blackbowtie}_{jk}\ =\  \sum_{i,\ell \in N} {A}^{\blackbowtie}_{(ij)(k
\ell)} \pr{r}^{\blackbowtie}_{i\ell}
\]
On the other hand, the node attraction $\pn{r}$ turns out to be a 
stationary distribution of a process that refines $\pr{A}^{\blackbowtie}$.

\begin{definition} Given a network $A$, its\/ {\em attraction dynamics} 
is a Markov chain $\pn{A} = \left(\pn{A}_{(ij)(k\ell)}\right)_{N^2\times 
N^2}$, with the entries
\bea
\pn{A}_{(ij)(k\ell)} & = & \frac{A_{ij} \pr{A}_{jk} A_{k\ell}}{A_{i \bullet } \pr
{A}_{\bullet \bullet } A_{\bullet\ell} } \label{detourfwd}
\eea
where $A_{i\bullet} \pr{A}_{\bullet\bullet} A_{\bullet\ell} = \sum_{m,n\in 
N} A_{im} \pr{A}_{mn} A_{n\ell}$.
\end{definition}

\begin{proposition}\label{propo}
Suppose that a given network $A$ is $v$-complete for a sufficiently 
large $v$. Then the node attraction $\pn{r}$, defined in (\ref{natt}), is 
the stationary distribution of its attraction dynamics (\ref{detourfwd}). In 
other words, for every $j,k$ holds
\bea
\pn{r}_{jk} & = & \sum_{i,\ell \in N} \pn{A}_{(ij)(k\ell)} \pn{r}_{i\ell}\label
{Rright}
\eea
\end{proposition}

The proof is in the Appendix. It is based on the following lemma.

\begin{lemma}\label{lemmapath} For a network $A$, which is $v$-
complete for a sufficiently large cutoff value $v$, the following 
equations hold for $i\stackrel{a}{\rightarrow}\ell$ and $j\stackrel{b}
{\rightarrow}k$
\bea 
\pn{A}_{\aaa\bbb} & = & \frac{A_\bbb}{4^d A_\aaa} A_{ij} A_{k\ell} 
\label{one}\\
\sum_{j\underset{c}{\rightarrow} k} \pn{A}_{\aaa c} & = &  \frac{1}{4^{d} 
A_\aaa} A_{ij} \pr{A}_{jk} A_{k\ell} \label{twojk}\\
\pn{A}_{\aaa\bullet} & = &  \frac{1}{4^{d} A_\aaa} A_{i\bullet} \pr{A}_
{\bullet \bullet} A_{\bullet \ell} \label{threealpha}
\eea
\end{lemma}

On the other hand, proposition \ref{propo} implies the following 
corollary, which establishes that formula (\ref{abias}) can be used to 
measure the attraction bias, as intended. 

\begin{corollary}\label{marginals}
The  directed reputation and promotion ranks are the marginals of the 
node attraction
\bea
\sum_{k\in N} \pn{r}_{jk} & = & r^\blacktriangleright_{j} \label{marg1} \\
\sum_{j\in N} \pn{r}_{jk} & = & r^\blacktriangleleft_{k} \label{marg2} 
\eea
\end{corollary}

\subsubsection{Interpretation.} 
To understand the meaning of attraction bias, consider a $v$-complete 
network $A$, with the forward-out and backward-in dynamics. The pull 
rank $r^\blacktriangleright_j$ tells how likely it is that a randomly 
sampled traffic path arrives to $j$; whereas the push rank $r^
\blacktriangleleft_k$ tells how likely it is that a randomly sampled traffic 
path departs from $k$. 

On the other hand, the attraction dynamics in the induced path 
network $\pn{A}$ gives the node attraction $\pn{r}_{jk}$, which tells 
how likely it is that a randomly sampled traffic path traverses a path 
from $j$ to $k$. In summary, we have
\bear 
r^\blacktriangleright_j  & =&  {\sf Prob}\big(\bullet\stackrel{\xi}
{\rightarrow} j \ |\ \xi\in A^\blacktriangleright\big)\\
r^\blacktriangleleft_k &  = & {\sf Prob}\big(k \stackrel{\xi}{\rightarrow}
\bullet \ |\ \xi\in A^\blacktriangleleft\big)\\
\pn{r}_{jk} &  = & {\sf Prob}\big( j \stackrel{\xi}{\rightarrow}k \ |\ \xi\in \pn
{A}\big)
\eear
Although the notation suggests that $r^\blacktriangleright$, $r^
\blacktriangleleft$, and $\pn{r}$ are sampled from different processes, 
corollary \ref{marginals} establishes that $\pn{r}$ is in fact the joint 
distribution of $r^\blacktriangleright$ and $r^\blacktriangleleft$. 

Nevertheless, a diligent reader will surely notice a twist of $j$ and $k$ 
in the last three equations, and wonder why is the probability that traffic goes from $j$ to $k$ related with the 
probabilities that it arrives {\em to\/} $j$, and that it departs {\em from\/} 
$k$? --- The answer to this question makes the forward-{\em out} and the backward-{\em in}\/ dynamics into a more interesting example than its many dynamical cousins. Briefly, if the surfers are more likely to flow with $\bullet{\rightarrow} j$ if the capacity of the links out of $j$ is higher, and if they are more likely to flow with $k {\rightarrow}\bullet$ if the capacity of the links into 
$k$ is higher, then the surfers are most likely to follow both these 
flows, i.e. into $j$ and out of $k$ --- if there is a high capacity of the 
links $j\rightarrow k$.

\subsubsection{Mutual information of the inputs and the outputs.} The 
fact that $\pn{r}$ is the joint distribution of the processes expressed by 
$r^\blacktriangleright$ and $r^\blacktriangleleft$ allows us to extract 
their {\em mutual information} \cite{Cover-Thomas}
\[
I(r^\blacktriangleright\ ;\ r^\blacktriangleleft)\  \ =\  \ D(\pn{r}\ ||\ r^
\blackbowtie) \ \  =\ \ \sum_{j=1}^N\sum_{k=1}^N \pn{r}_{jk}\ \log \frac
{\pn{r}_{jk}}{r^\blacktriangleright_j r^\blacktriangleleft_k}
\]
Its expression in terms of relative entropy $D(\pn{r}\ ||\ r^\blackbowtie)
$ [{\em ibidem}] shows that it measures how much we lose, in the 
efficiency of encoding of $\pn{r}$ if we assume that $r^
\blacktriangleright$ and $r^\blacktriangleleft$ are mutually 
independent. Intuitively, the mutual information $I(r^\blacktriangleright\ 
;\ r^\blacktriangleleft)$ can thus be taken as a measure of the {\em 
locality\/} of information processing in $A$. If this is an entirely local 
process, then every path must begin and end at the same node, and 
the random walks $\delta$ and $\varrho$, selecting the sources and 
the destinations of the paths, must coincide. But if $\delta = \varrho$, 
then the push rank and the pull rank must obey the same distribution 
$r^\blacktriangleleft = r^\blacktriangleright = r$, and their mutual 
information is $I(r^\blacktriangleright\ ;\ r^\blacktriangleleft) = H(r)$, 
their entropy. In the other extreme case, the random walks $\delta$ 
and $\varrho$ are independent\footnote{The theorem in the appendix 
suggests that they are similarly distributed, up to a scale factor.}, and 
their joint distribution is just the product of their distributions $\pn{r}_
{jk} =  r^\blacktriangleright_j r^\blacktriangleleft_k$.  Their mutual 
information is then $I(r^\blacktriangleright\ ;\ r^\blacktriangleleft) = 0$.

\section{Conclusions and future work}
When the Web is viewed as a global data store, the problem of its 
semantics is the problem of determining a uniform meaning for the 
data published by its various participants. The search engines are dealing with this problem on the level of the human-Web interaction (e.g., distinguishing the meanings of the word "jaguar", sometimes denoting a car, sometimes an animal \cite{kleinberg99authoritative}, or deciding whether "Paris Hilton", in a given context, refers to a person or  to a hotel, etc.), whereas the Semantic Web project \cite{SemanticWeb} deals with the computer-Web interactions. 
When the Web is viewed as a computer, the problem of its 
semantics is not just a matter of assigning some meanings to some data stored in it, but also to its data processing operations. For programming languages, this is what we usually call operational semantics. However, unlike a programming language, the Web, and other spontaneously evolving networks, do not have a formally defined set of data structures and operations: data are transformed by many random walks, running concurrently. Operational semantics of network computation requires a toolkit for incremental analysis of such processes. In this paper, we described a path ranking method, which is may become a useful piece of that toolkit. Now we sketch a way to test it experimentally. Using the notion of attraction bias, we lift the graph theoretic notion of (maximal) {\em clique\/} into rank analysis, while retaining network dynamics as a graph structure over such generalized cliques. We call these generalized cliques {\em concepts\/} and the links between them {\em associations}. 

\subsubsection*{Communities and concepts.}%\label{Concepts}
Taking the notion of attraction bias back to the idea of communities as 
sets of nodes with high cohesion, from which we started in the 
Introduction, we now reformulate the notion of cohesion in a 
different norm ($\ell_\infty$ instead of $\ell_1$), and define cohesion of 
a set of nodes $U\subseteq N$ to be their minimal symmetric 
attraction bias
\bear
\bias(U) & = & \bigwedge_{i,j \in U} (\bias_{ij} \vee \bias_{ji})
\eear
For each $\varepsilon\in [0,1]$, we define an {\em $\varepsilon$-
community\/} to be a set of nodes $U\subseteq N$ such that $\bias(U)
\geq \varepsilon$. Denoting by $\wp_\varepsilon N$ the set of $
\varepsilon$-communities, note that $\varepsilon_1\leq \varepsilon_2$ 
implies that $\wp_{\varepsilon_1}N \supseteq \wp_{\varepsilon_2} N$.
The partial ordering of $U,V\in \wp_\varepsilon N$ is given by 
$
U\sqsubseteq V\ \iff \ U\subseteq V \wedge \bias(U)\leq \bias(V)
$
This gives a {\em directed complete partial order (dcpo)}. 
It is not a lattice because some communities cannot be extended by 
new nodes without decreasing their cohesion; so there are pairs of 
communities that cannot be joined, and do not have an upper bound. 
However, {\em directed\/} sets of communities (i.e., where each pair 
has an upper bound) do have least upper bounds, which are just their 
set theoretic unions. Directed complete partial orders are often used in  
denotational semantics of programming languages \cite
{Compendium}. According to that interpretation, communities can be 
thought of as pieces of {\em partial information}, their $\sqsubseteq$-
ordering as the increase of information, and the existence of an upper 
bound of two communities as the {\em consistency\/} of the 
informations that they carry. 

The maximal elements of $\wp_\varepsilon N$, i.e. the communities 
that cannot be extended by new nodes without losing cohesion, can 
be construed as {\em $\varepsilon$-concepts}. A set $U\in \wp_
\varepsilon N$ is thus an $\varepsilon$-concept if $\bias\left(\{i,j\}\right)
\geq \varepsilon$ holds for all $i,j\in U$, but for every $k\in N\setminus 
U$ there is a $j\in U$ such that $\bias\left(\{k,j\}\right)\lt \varepsilon$.

The community and concept structure of a network $A$ can be 
analyzed by studying the sequence of hypergraphs $A_\varepsilon$, 
where the $\varepsilon$-concepts, or the $\varepsilon$-communities 
approximating them, are viewed as hyperedges. The sequence $(A_
\varepsilon)_{\varepsilon\in [0,1]}$ decreases as the cohesion 
parameter $\varepsilon$ increases, and the highly cohesive 
communities and concepts can be feasibly analyzed.

A level further, concepts and communities can be viewed as the nodes 
of a network. The most interesting definition of the links between them, 
intuitively thought of as associations, is based on a variant of a path 
network, complementing definition \ref{def2}. A sketch of this definition 
is in the next, final subsection.

\subsubsection*{Associations.}
Let $\NNN^\varepsilon$ denote the set of $\varepsilon$-concepts in a 
network $A$. 

The {\em concept network\/} $\AAA^\varepsilon$, induced by a 
network $A$, has the $\varepsilon$-concepts as its nodes. Its edges 
are called {\em concept associations}. The set of associations 
between $U,V\in \NNN^\varepsilon$  is
\bea\label{association}
\EEE^\varepsilon_{UV} & = &  \sum_{U\underset{\aaa}{\rightarrow}U
\cap V}\ \ \sum_{U\cap V\underset{\bbb}{\rightarrow}V}\  \cn{E}_{\aaa
\bbb}
\eea
where $U\stackrel{\xi}{\rightarrow} V$ abbreviates $\delta(\xi)\in U\ 
\wedge\ \varrho(\xi)\in V$, and
\bear
\cn{E}_{\aaa\bbb} & = &  \big\{f=<f_0,f_1>\in E_{ij}\times E_{k\ell}\ |\  
 \gamma(f_0)+ \gamma(\bbb) \leq v-d 
\mbox{ and } \gamma(\aaa) + \gamma(f_1) \leq v-d \big\}
\eear
An association $f \in \AAA_{UV}$ is thus a quadruple $f = <\aaa,
\bbb,f_0,f_1>$
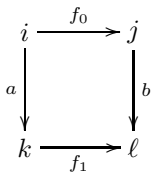
\[
\xymatrix@C-1pc@R-1.5pc
{ i \ar[ddd]_\aaa 
\ar[rr]^{f_0} && j \ar[ddd]^\bbb  \\  \\ \\
k \ar[rr]_{f_1} && \ell 
}
\]
such that $i,j,k\in U$ and $j,k,\ell\in V$. Its cost is 
$
\gamma(f) \ = \ \gamma(f_0) + \gamma(\bbb) - \gamma(\aaa)- 
\gamma(f_1) 
$. 
The cost of an association from $U$ to $V$ is lower if the traffic from 
$i\in U$ to $\ell \in V$ gets less costly when it crosses to $V$ earlier.

While the general network analysis tools apply to concept networks, 
the various notions of dynamics acquire new meanings on this level. At 
this point, understanding which of the possible interpretations may 
lead to useful tools for extracting and analyzing the relevant concepts, 
processed in a network, seems to call for experimentation with real 
data.

\bibliographystyle{abbrv}
\bibliography{scale}

\pagebreak
\appendix
\section*{Appendix: Proofs}
\begin{proof}[of lemma \ref{lemmapath}(\ref{one})]
The first claim is that there is a sufficiently large $v$ such that $
\gamma(f_0)+ \gamma(\bbb)+ \gamma(f_1) \leq v-2d$ hods for all 
$f_0\in E_{ij}$ and $f_1\in E_{k\ell}$. Since $\bbb$ and $d$ are fixed, 
the claim is clear if $E_{ij}$ and $E_{k\ell}$ are finite. Since $A$ is 
assumed to be truncated complete, an infinite set of paths can only be 
generated from the links with a cost $\leq 0$. So the costs of the 
elements of $E_{ij}$ and $E_{k\ell}$ are in any case bounded.

But if all $f_0\in E_{ij}$ and $f_1\in E_{k\ell}$ satisfy $\gamma(f_0)+ 
\gamma(\bbb)+ \gamma(f_1) \leq v-2d$, then $\pn{E}_{\aaa\bbb}  =   
E_{ij}\times E_{k\ell}$. Unfolding the definition of $\pn{A}_{\aaa\bbb}$ 
and using (\ref{gamma}) we get 
\bear
\pn{A}_{\aaa\bbb} & = & \sum_{f\in \pn{E}_{\aaa\bbb}} 2^{-\gamma(f)}\\
& = & \sum_{f_0 \in E_{ij}}\sum_{f_1\in E_{k\ell}} 2^{-2d -\gamma(f_0)-
\gamma(\bbb)-\gamma(f_1)+\gamma(\aaa)}\\
& = & 4^{-d}\ \frac{2^{-\gamma(\bbb)}}{2^{-\gamma(\aaa)}} \sum_{f_0
\in E_{ij}} 2^{-\gamma(f_0)}\sum_{f_1\in E_{k\ell}} 2^{-\gamma(f_1)}\\
 & = &\frac{A_\bbb}{4^{d} A_\aaa} A_{ij} A_{k\ell}
 \eear
\ref{lemmapath}(\ref{twojk}) follows directly from \ref{lemmapath}(\ref
{one}), unpacking $A_e = 2^{-\gamma(e)}$. And \ref{lemmapath}(\ref
{threealpha}) then follows from \ref{lemmapath}(\ref{twojk}):
\bear
\pn{A}_{\aaa\bullet} & = & \sum_{\bbb} \pn{A}_{\aaa\bbb} \\
& = & \sum_{j,k\in N} \sum_{j\underset{\bbb}{\rightarrow} k} \pn{A}_
{\aaa\bbb} \\
& = & \sum_{j,k\in N} \frac{1}{4^{d}A_\aaa} A_{ij} \pr{A}_{jk} A_{k\ell} \\
& = & \frac{1}{4^{d}A_\aaa} A_{i\bullet} \pr{A}_{\bullet \bullet} A_{\bullet 
\ell}
\eear
\end{proof}

\begin{proof}[of proposition \ref{propo}(\ref{Rright})] 
We unfold the definition of $\pn{r}$ and then use (\ref{twojk}) and (\ref
{threealpha}):
\bear
\pn{r}_{jk} & = & \sum_{j\underset{\bbb}{\rightarrow}k} \pn{r}_\bbb \\
& = & \sum_{j\underset{\bbb}{\rightarrow}k } \sum_{\aaa} \frac{\pn{A}_
{\aaa\bbb}}{\pn{A}_{\aaa \bullet}} \pn{r}_\aaa \\
& = & \sum_{\aaa} \frac{\sum_{j\underset{\bbb}{\rightarrow}k}  \pn{A}_
{\aaa\bbb}}{\pn{A}_{\aaa\bullet}} \pn{r}_\aaa\\
& = & \sum_{\aaa} \frac{\frac{1}{4^{d}A_\aaa}A_{ij} \pr{A}_{jk} A_{k\ell}}
{\frac{1}{4^{d}A_\aaa} A_{i\bullet} \pr{A}_{\bullet \bullet} A_{\bullet \ell}} 
\pn{r}_\aaa\\
& = & \sum_{i,\ell \in N} \sum_{i\underset{\aaa}\rightarrow \ell} \frac{A_
{ij} \pr{A}_{jk} A_{k\ell}}{ A_{i\bullet} \pr{A}_{\bullet \bullet} A_{\bullet 
\ell}} \pn{r}_\aaa\\
& = & \sum_{i,\ell \in N}  \frac{A_{ij} \pr{A}_{jk} A_{k\ell}}{ A_{i\bullet} \pr
{A}_{\bullet \bullet} A_{\bullet \ell}} \sum_{i\underset{\aaa}\rightarrow 
\ell} \pn{r}_\aaa \\
& = &  \sum_{i,\ell \in N}  \pn{A}_{(ij)(k\ell)} \pn{r}_{i\ell}
\eear
\end{proof}

\begin{proof}[of corollary \ref{marginals}(\ref{marg1})] We set $q_j = 
\sum_{k\in N} \pn{r}_{jk}$ and expand $\pn{r}_{jk}$ using proposition 
\ref{propo}(\ref{Rright}), to show that $q$ is the stationary distribution 
of the process $\pr{A^\triangleright}$:
\bear
q_j & = & \sum_{k\in N} \pn{r}_{jk}\\
 & = & \sum_{k,i,\ell \in N} \pn{A}_{(ij)(k\ell)} \pn{r}_{i\ell} \\
& = & \sum_{k,i,\ell \in N} \frac{A_{ij} \pr{A}_{jk} A_{k\ell}}{A_{i \bullet } 
\pr{A}_{\bullet \bullet } A_{\bullet\ell} } \pn{r}_{i\ell}\\
& = &  \sum_{i,\ell  \in N} \frac{A_{ij} \pr{A}_{j\bullet }}{A_{i \bullet } \pr
{A}_{\bullet \bullet }} \pn{r}_{i\ell} \\
& = &  \sum_{i \in N} A^\triangleright_{ij}\  \frac{\pr{A}_{j\bullet }}{\pr{A}_
{\bullet \bullet }}  \sum_{\ell \in N} \pn{r}_{i\ell} \\
& = & \sum_{i \in N} \pr{A}^\blacktriangleright_{ij}  q_i
\eear
Since $\pr{r^\blacktriangleright}$ is by definition the stationary point of $\pr{A^\blacktriangleright}$, the uniqueness implies $q = \pr{r^\blacktriangleright}$. 

To prove \ref{marginals}(\ref{marg2}), we set $q_k = \sum_{j\in N} \pn
{r}_{jk}$ and proceed similarly:
\bear
q_k & = & \sum_{j\in N} \pn{r}_{jk}\\
 & = & \sum_{j,i,\ell \in N} \pn{A}_{(ij)(k\ell)} \pn{r}_{i\ell} \\
& = & \sum_{j,i,\ell \in N} \frac{A_{ij} \pr{A}_{jk} A_{k\ell}}{A_{i \bullet } 
\pr{A}_{\bullet \bullet } A_{\bullet\ell} } \pn{r}_{i\ell}\\
& = & \sum_{i,\ell \in N} \frac{\pr{A}_{\bullet k} A_{k\ell}}{\pr{A}_{\bullet 
\bullet } A_{\bullet\ell} } \pn{r}_{i\ell}\\
& = &  \sum_{\ell \in N} \frac{\pr{A}_{\bullet k}}{\pr{A}_{\bullet \bullet }}\  
A^\triangleleft_{k\ell }  \sum_{i \in N} \pn{r}_{i\ell} \\
& = & \sum_{\ell \in N} \pr{A}^\blacktriangleleft_{k\ell}  q_\ell
\eear
Since $\pr{r^\blacktriangleleft}$ is by definition the stationary point of $\pr{A^\blacktriangleleft}$, the uniqueness implies $q = \pr{r^\blacktriangleleft}$.

\end{proof}

\end{document}